\documentclass{ws-procs975x65}
\usepackage{color}

\begin{document}


\wstoc{Topological Gravitation on Graph Manifolds}{N.V.
Mitskievich}

\title{TOPOLOGICAL GRAVITATION ON GRAPH MANIFOLDS}

\author{N.V. MITSKIEVICH, V.N. EFREMOV, AND A.M. HERN\'ANDEZ
MAGDALENO}
\address{CUCEI, Universidad de Guadalajara,
Guadalajara, Jalisco, Mexico\\
Apdo. Postal 1-2011 C.P. 44100, Guadalajara, Jalisco, M\'exico\\
\email{mitskievich03@yahoo.com.mx}}

\begin{abstract}
A model of topological field theory is presented in which the
vacuum coupling constants are topological invariants of the
four-dimensional spacetime. Thus the coupling constants are
theoretically computable, and they indicate the topological
structure of our universe.
\end{abstract}

\bodymatter

We construct an Abelian $BF$-type model in analogy with the
ordinary four-dimensional topological field theory \cite{Thom} and
with the low-energy effective $U(1)^r$-theory of Seiberg--Witten
(SW)\cite{Arg} , beginning with a $U(1)^r$-bundle $E$ over a
four-dimensional topological space $X$ with a non-empty boundary
$\partial X$, $E$ being a direct sum of linear bundles
$L_1\oplus\cdots\oplus L_r$. Let us define locally connection
1-forms $A^a$ ($a=1,\dots,r$) on $E$ with values in the algebra
$L$ of the group $U(1)$, and 2-forms $B_a$ with values in the dual
algebra. Due to these analogies it is natural to write the action
as $S=\int{F^a\wedge B_a-\frac{1}{2}\Lambda^{ab}B_a\wedge
B_b+\frac{i}{2}\Theta_{ab}F^a\wedge F^b}$ where $F^a=dA^a$;
$\Lambda^{ab}$ and $\Theta_{ab}$ are non-degenerate symmetric
matrices called those of the coupling constants and theta angles
matrices, respectively. Our action admits symmetry under dual
conjugation similar to the electro--magnetic one (EM duality) of
the SW theory, namely $ F^a\rightarrow F_{Da}=\Lambda_{ab}F^b; ~~
B_a\rightarrow B_D^a=\Lambda^{ab}B_b$; $\Lambda^{ab}\rightarrow
\Lambda_{ab}; ~~
\Theta_{ab}\rightarrow\Theta^{ab}=\Lambda^{ac}\Lambda^{bd}
\Theta_{cd}, ~~ \Lambda_{ab}\Lambda^{bc}=\delta_a^c. $ These
duality transformations carry $S$ into its equivalent, $S_D$, and
they involve the strong--weak coupling duality.\cite{Arg} We call
$\Lambda_{ab}$ and $\Lambda^{ab}$ strong and weak coupling
constants matrices (in this sense), respectively.

\begin{figure}[t]
\begin{center}
 \vspace*{1.7cm} 
 \setlength{\unitlength}{.65pt}
\begin{picture}(190,190)
\put(-19,255){\line(0,-1){30}} \put(-20,255){\line(1,-1){30}}
\put(-20,255){\line(1,0){30}} \put(236,191){\line(0,-1){29}}
\put(236,191){\line(1,0){30.5}} \put(236,191){\line(-1,0){30}}
\put(236,126){\line(0,-1){29}} \put(236,126){\line(1,0){30.5}}
\put(236,126){\line(-1,0){30}} \put(236,64){\line(0,-1){30}}
\put(236,64){\line(1,0){30.5}} \put(236,64){\line(-1,0){30}}
\put(236,255){\line(0,-1){30}} \put(236,255){\line(1,0){30.5}}
\put(236,255){\line(-1,0){30}} \put(172,126){\line(0,-1){29}}
\put(172,126){\line(1,0){30}} \put(172,126){\line(-1,0){30}}
\put(172,191){\line(0,-1){30}} \put(172,191){\line(1,0){30}}
\put(172,191){\line(-1,0){30}} \put(172,255){\line(0,-1){30}}
\put(172,255){\line(1,0){30}} \put(172,255){\line(-1,0){30}}
\put(107,191){\line(0,-1){30}} \put(107,191){\line(1,0){30}}
\put(107,191){\line(-1,0){30}} \put(107,255){\line(0,-1){30}}
\put(107,255){\line(1,0){30}} \put(107,255){\line(-1,0){29}}
\put(45,255){\line(0,-1){30}} \put(44,255){\line(1,0){30}}
\put(44,255){\line(-1,0){30}} \put(235,00){\line(0,-1){30}}
\put(235,00){\line(1,0){32}} \put(235,00){\line(-1,1){29}}
\put(43,191){\line(1,0){30}} \put(43,191){\line(-1,1){29}}
\put(43,191){\line(1,-1){30}} \put(108,126){\line(1,0){30}}
\put(108,126){\line(-1,1){29}} \put(108,126){\line(1,-1){29}}
\put(170,64){\line(1,0){30}} \put(170,64){\line(-1,1){29}}
\put(170,64){\line(1,-1){30}}
\put(236,225){\makebox(.5,.5){$\bullet$}}
\put(172,225){\makebox(.5,.5){$\bullet$}}
\put(107,225){\makebox(.5,.5){$\bullet$}}
\put(45,225){\makebox(.5,.5){$\bullet$}}
\put(236,35){\makebox(.5,.5){$\bullet$}}
\put(172,161){\makebox(.5,.5){$\bullet$}}
\put(107,161){\makebox(.5,.5){$\bullet$}}
\put(236,96){\makebox(.5,.5){$\bullet$}}
\put(172,96){\makebox(.5,.5){$\bullet$}}
\put(236,161){\makebox(.5,.5){$\bullet$}}
\put(235,-30){\makebox(.5,.5){$\bullet$}}
\put(268,-0.5){\makebox(.5,.5){$\bullet$}}
\put(-19,224){\makebox(.5,.5){$\bullet$}}
\put(268,63){\makebox(.5,.5){$\bullet$}}
\put(268,125.5){\makebox(.5,.5){$\bullet$}}
\put(268,190){\makebox(.5,.5){$\bullet$}}
\put(268,254.5){\makebox(.5,.5){$\bullet$}}
\put(206.5,28){\makebox(.5,.5){$\bullet$}}
\put(200.5,33){\makebox(.5,.5){$\bullet$}}
\put(141.5,92){\makebox(.5,.5){$\bullet$}}
\put(136,97){\makebox(.5,.5){$\bullet$}}
\put(79,154){\makebox(.5,.5){$\bullet$}}
\put(74,159){\makebox(.5,.5){$\bullet$}}
\put(14.5,219){\makebox(.5,.5){$\bullet$}}
\put(10.5,223.5){\makebox(.5,.5){$\bullet$}}
\put(201,254){\makebox(.5,.5){$\bullet$}}
\put(207,254){\makebox(.5,.5){$\bullet$}}
\put(136,254){\makebox(.5,.5){$\bullet$}}
\put(142,254){\makebox(.5,.5){$\bullet$}}
\put(73,254){\makebox(.5,.5){$\bullet$}}
\put(79,254){\makebox(.5,.5){$\bullet$}}
\put(9,254){\makebox(.5,.5){$\bullet$}}
\put(15,254){\makebox(.5,.5){$\bullet$}}
\put(201,63){\makebox(.5,.5){$\bullet$}}
\put(207,63){\makebox(.5,.5){$\bullet$}}
\put(136,190){\makebox(.5,.5){$\bullet$}}
\put(142,190){\makebox(.5,.5){$\bullet$}}
\put(72,190){\makebox(.5,.5){$\bullet$}}
\put(78,190){\makebox(.5,.5){$\bullet$}}
\put(201,125){\makebox(.5,.5){$\bullet$}}
\put(207,125){\makebox(.5,.5){$\bullet$}}
\put(136,125){\makebox(.5,.5){$\bullet$}}
\put(142,125){\makebox(.5,.5){$\bullet$}}
\put(201,190){\makebox(.5,.5){$\bullet$}}
\put(207,190){\makebox(.5,.5){$\bullet$}}
\put(236,225){\circle{10}} \put(172,225){\circle{10}}
\put(107,225){\circle{10}} \put(45,225){\circle{10}}
\put(236,35.5){\circle{10}} \put(172,161.5){\circle{10}}
\put(107,161.5){\circle{10}} \put(236,96.5){\circle{10}}
\put(172,96.5){\circle{10}} \put(236,161.5){\circle{10}}
\put(235,-29){\circle{10}} \put(268,00){\circle{10}}
\put(-18.5,224){\circle{10}} 
\put(268.5,64){\circle{10}} \put(268,126){\circle{10}}
\put(268,190.5){\circle{10}} \put(268,255){\circle{10}}
\put(204,255){\oval(20,10)} \put(139,255){\oval(20,10)}
\put(76,255){\oval(20,10)} \put(12,255){\oval(20,10)}
\put(204,64){\oval(20,10)} \put(139,191){\oval(20,10)}
\put(75,191){\oval(20,10)} \put(204,126){\oval(20,10)}
\put(139,126){\oval(20,10)} \put(204,191){\oval(20,10)}
\put(9,220){\qbezier(7.9497, 6.9497)(3.8492, 11.0503)(-0.5, 12.5)
\qbezier(-0.5, 12.5)(-4.8492, 13.9497)(-6.8995, 11.8995)
\qbezier(-6.8995, 11.8995)(-8.9497, 9.8492)(-7.5, 5.5)
\qbezier(-7.5, 5.5)(-6.0503, 1.1508)(-1.9497, -2.9497)
\qbezier(-1.9497, -2.9497)(2.1508, -7.0503)(6.5, -8.5)
\qbezier(6.5, -8.5)(10.8492, -9.9497)(12.8995, -7.8995)
\qbezier(12.8995, -7.8995)(14.9497, -5.8492)(13.5, -1.5)
\qbezier(13.5, -1.5)(12.0503, 2.8492)(7.9497, 6.9497)}
\put(73,155){\qbezier(7.9497, 6.9497)(3.8492, 11.0503)(-0.5, 12.5)
\qbezier(-0.5, 12.5)(-4.8492, 13.9497)(-6.8995, 11.8995)
\qbezier(-6.8995, 11.8995)(-8.9497, 9.8492)(-7.5, 5.5)
\qbezier(-7.5, 5.5)(-6.0503, 1.1508)(-1.9497, -2.9497)
\qbezier(-1.9497, -2.9497)(2.1508, -7.0503)(6.5, -8.5)
\qbezier(6.5, -8.5)(10.8492, -9.9497)(12.8995, -7.8995)
\qbezier(12.8995, -7.8995)(14.9497, -5.8492)(13.5, -1.5)
\qbezier(13.5, -1.5)(12.0503, 2.8492)(7.9497, 6.9497)}
\put(136,93){\qbezier(7.9497, 6.9497)(3.8492, 11.0503)(-0.5, 12.5)
\qbezier(-0.5, 12.5)(-4.8492, 13.9497)(-6.8995, 11.8995)
\qbezier(-6.8995, 11.8995)(-8.9497, 9.8492)(-7.5, 5.5)
\qbezier(-7.5, 5.5)(-6.0503, 1.1508)(-1.9497, -2.9497)
\qbezier(-1.9497, -2.9497)(2.1508, -7.0503)(6.5, -8.5)
\qbezier(6.5, -8.5)(10.8492, -9.9497)(12.8995, -7.8995)
\qbezier(12.8995, -7.8995)(14.9497, -5.8492)(13.5, -1.5)
\qbezier(13.5, -1.5)(12.0503, 2.8492)(7.9497, 6.9497)}
\put(200,29){\qbezier(7.9497, 6.9497)(3.8492, 11.0503)(-0.5, 12.5)
\qbezier(-0.5, 12.5)(-4.8492, 13.9497)(-6.8995, 11.8995)
\qbezier(-6.8995, 11.8995)(-8.9497, 9.8492)(-7.5, 5.5)
\qbezier(-7.5, 5.5)(-6.0503, 1.1508)(-1.9497, -2.9497)
\qbezier(-1.9497, -2.9497)(2.1508, -7.0503)(6.5, -8.5)
\qbezier(6.5, -8.5)(10.8492, -9.9497)(12.8995, -7.8995)
\qbezier(12.8995, -7.8995)(14.9497, -5.8492)(13.5, -1.5)
\qbezier(13.5, -1.5)(12.0503, 2.8492)(7.9497, 6.9497)}
\put(-19,254){\makebox(.5,.5){$\bullet$}}
\put(44,190){\makebox(.5,.5){$\bullet$}}
\put(109,125){\makebox(.5,.5){$\bullet$}}
\put(172,190){\makebox(.5,.5){$\bullet$}}
\put(236,254){\makebox(.5,.5){$\bullet$}}
\put(45,254){\makebox(.5,.5){$\bullet$}}
\put(107,254){\makebox(.5,.5){$\bullet$}}
\put(172,254){\makebox(.5,.5){$\bullet$}}
\put(235,0){\makebox(.5,.5){$\bullet$}}
\put(107,190){\makebox(.5,.5){$\bullet$}}
\put(171,63){\makebox(.5,.5){$\bullet$}}
\put(236,63){\makebox(.5,.5){$\bullet$}}
\put(172,125){\makebox(.5,.5){$\bullet$}}
\put(236,125){\makebox(.5,.5){$\bullet$}}
\put(236,190){\makebox(.5,.5){$\bullet$}}
\put(204,0){\makebox(.5,.5){${\Sigma}${\scriptsize(1,1,2)}}}
\put(141,61){\makebox(.5,.5){${\Sigma}${\scriptsize(2,3,5)}}}
\put(73,124){\makebox(.5,.5){${\Sigma}${\scriptsize(30,7,11)}}}
\put(-2,188){\makebox(.5,.5){${\Sigma}${\scriptsize(2310,13,17)}}}
\put(-70,253){\makebox(.5,.5){${\Sigma}${\scriptsize(510510,19,23)}}}
\end{picture}
\end{center}
\vspace*{20.pt}\caption{The graph associated with a
four-dimensional manifold interpreted as a Euclidean region of
spacetime\! ${X}$ with boundaries $\partial{X}=-\bigsqcup_{ i=1}^I
L(p_i,q_i)\bigsqcup M$. This graph describes a universe with five
low energy interactions related to the first nine prime numbers as
(1,2), (3,5), (7,11), (13,17), (19,23) determining Seifert's
invariants of Sfh-spheres being glued together. Vertices on
horizontal lines represent (from left to right) successive
``derivatives'' of the primary Sfh-spheres shown on the diagonal
line. At the four-dimensional level splicing gives the plumbing
operation.} \label{fig1}
\end{figure}
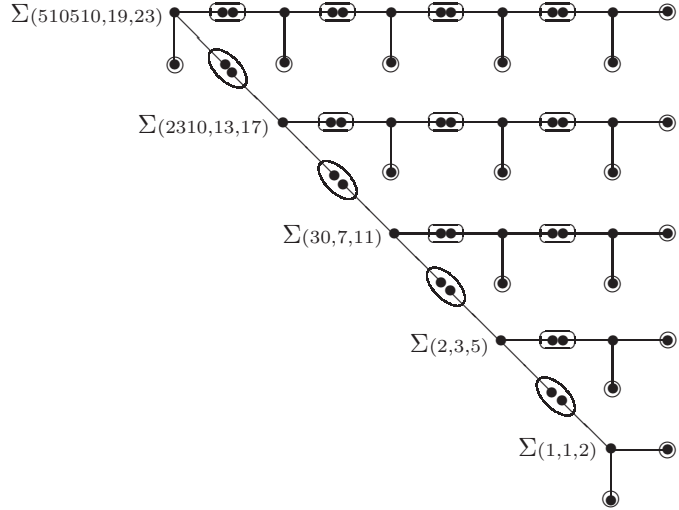

Then we generalize Dirac's quantization conditions: the flux
through non-trivial 2-cycles $\Sigma_I$ must be
$\int_{\Sigma_I}F^a=2\pi m^a_I, ~~ m^a_I\in \mathbb{Z}.$ Then from
dynamical equations of our $BF$ system $~ dB_a=0, ~~
F^a=\Lambda^{ab}B_b $ together with the gauge symmetries it
follows that the moduli space of this $BF$ system is $H^2(X,
\mathbb{Z})\oplus H^2(X,\partial X, \mathbb{Z})$ meaning that
$[\frac{1}{2\pi} F^a] \in H^2(X, \mathbb{Z})$ and $[\frac{1}{2\pi}
B_a] \in H^2(X,\partial X, \mathbb{Z}).$ Thus there are no local
degrees of freedom and like in the case of the low energy
effective SW action, our model describes the moduli space of
vacua. The spacetime topology is non-trivial since we model the
spacetime by the graph manifold \cite{EM}. Each tree graph
corresponds to a unique four-dimensional space $X$ with a boundary
containing lens spaces and $\mathbb{Z}$-homology spheres. The
latter ones are results of splicing of Seifert fibred homology
(Sfh) spheres. The most important construction element is
Sfh-sphere $\Sigma (\underline{a})\equiv\Sigma(a_1,a_2,a_3)$
having three special orbits and being a three-dimensional manifold
which is an intersection of the Brieskorn surface
${z_1}^{a_1}+{z_2}^{a_2} +{z_3}^{a_3}=0$ ($z_i \in \mathbb{C}_i$)
and a sphere $S^5$. Here $a_1$, $a_2$ and $a_3$ are mutually prime
integers (Seifert invariants). To the end of constructing our
cosmological model we need a specific family of Sfh-spheres to
which we give the following definition \cite{EM}: We take a
succession of Sfh-spheres calling it the primary one:
$\{\Sigma(q_{2n-1},p_{2n},p_{2n+1})|n=0,\dots ,4\}$. Here $p_i$ is
the $i$-th prime number in the natural series and $q_i=p_1\cdots
p_i$. Then we define the ``derivative'' of a Sfh-sphere $\Sigma
(\underline{a})$ as another Sfh-sphere
$\Sigma^{(1)}(\underline{a}):=\Sigma (a_1,a_2a_3,a+1)\equiv \Sigma
(a^{(1)}_1,a^{(1)}_2,a^{(1)}_3)$.The spatial sections $M$ of our
universe model we construct gluing together (by splicing) the
Sfh-spheres in agreement with tree-type graphs like that we give
in Fig. \ref{fig1}.

It is interesting to compute the intersection matrices for these
manifolds. But let us first recall that due to the
Poincar\'e--Lefschetz duality $\omega_\mathbb{Z}:H^2(X,\partial X,
\mathbb{Z})\otimes H^2(X, \mathbb{Z})\rightarrow\mathbb{Z}$ the
integer intersection form can be determined as a cup product
$\omega_\mathbb{Z}(b,f)=\left<b\cup f,[X,\partial X]\right>$ for
any $b\in H^2(X,\partial X, \mathbb{Z})$ and $f\in
H^2(X,\mathbb{Z})$. This is a generalization of the intersection
form in the de Rham representation, $\int_X b\wedge f.$ It is well
known that, if one takes some bases $b_i$ and $f^i$ of groups
$H^2(X,\partial X,\mathbb{Z})$ and $H^2(X,\mathbb{Z})$
($i=1,\dots,r, ~ r={\sf rank} H^2(X,\mathbb{Z})$) which are
mutually dual in the sense that $\omega_\mathbb{Z}
(b_i,f^j)=\delta_i^j$, the integer intersection matrix
$\omega_\mathbb{Z}(b_i,b_j)$ will be inverse to the rational one
$\omega_\mathbb{Q}(f^i,f^j)$. These  intersection matrices
represent basic topological invariants of any graph manifold. We
consider solutions of the dynamical equations $dB^a=0$,
$F^a=\Lambda^{ab}B_b$ as mutually dual bases of the respective
cohomology groups, so that $f^a=[\frac{1}{2\pi} F^a]$ and
$b_a=[\frac{1}{2\pi} B_a]$. The second equation yields $
\omega_\mathbb{Q}(f^a,f^b)=\Lambda^{ab},$
$\omega_\mathbb{Z}(b_a,b_b)=\Lambda_{ab}$. Thus the coupling
constants matrices $\Lambda^{ab}$ and $\Lambda_{ab}$ should be
identified with the rational and integer intersection matrices,
respectively. Now we return to the graph which describes spacetime
of our model. Each free vertex of this graph (i.e., each
Sfh-sphere) corresponds to an element $b_a$ of the basis and to an
element $f^a$ of the dual basis. Then the intersection matrix
elements correspond to the respective Sfh-spheres forming the
graph. We associate with any Sfh-sphere a certain interaction
(inclusion of one more Sfh-sphere into our universe model results
in switching on one new interaction). We associate with the
Sfh-spheres in the extreme right-hand ``column'' low-energy (LE)
interactions (strong, electromagnetic, weak, gravitational and
cosmological). Elimination of these Sfh-spheres yields a graph
describing the earlier stage of the cosmological evolution and a
family of higher energy interactions, thus our model involves an
interactions unification scheme. Each Sfh-sphere has three edges
corresponding to three special orbits, so one may glue it into the
graph in three different manners which yields $3^{15}$ different
universes. A hypothesis that the ``real universe'' is found in a
mixed state suggests that the physical meaning belongs to the
average intersection matrix:

\noindent\vspace*{7.pt}\hspace*{-3.5pt} {\footnotesize $\left(
\begin{array}{lllllllllllllll}
10^{\text{-4}} & 0 & 0 & 0 & {\color{red}\mathbf{10^\textbf{-3}}}
& 0 & 0 & 0
& 0 & 0 & 0 & 0 & 0 & 0 & 0 \\
0 & 10^{\text{-20}} & {\color{red}\mathbf{10^\textbf{-13}}} & 0 &
0 & 0 & 0 &
0 & 0 & 0 & 0 & 0 & 0 & 0 & 0 \\
0 & {\color{red}\mathbf{10^\textbf{-13}}} & 10^{\text{-6}} &
10^{\text{-7}} &
0 & 0 & 0 & 0 & 0 & 0 & 0 & 0 & 0 & 0 & 0 \\
0 & 0 & 10^{\text{-7}} & 10^{\text{-4}} & 10^{\text{-4}} & 0
& 10^{\text{-6}} & 0 & 0 & 0 & 0 & 0 & 0 & 0 & 0 \\
{\color{red}\mathbf{10^\textbf{-3}}} & 0 & 0 & 10^{\text{-4}} &
10^{\text{-2}}
& {\color{red}\mathbf{10^{\textbf{-2}}}} & 0 & 0 & 0 & 0 & 0 & 0 & 0 & 0 & 0 \\
0 & 0 & 0 & 0 & {\color{red}\mathbf{10^\textbf{-2}}} &
10^{\text{-1}} & 0 & 0
& 0 & 0 & 0 & 0 & 0 & 0 & 0 \\
0 & 0 & 0 & 10^{\text{-6}} & 0 & 0 & 10^{\text{-5}} &
10^{\text{-12}} & 0 & 0 & 10^{\text{-10}} & 0 & 0 & 0 & 0 \\
0 & 0 & 0 & 0 & 0 & 0 & 10^{\text{-12}} & 10^{\text{-11}}
& 10^{\text{-23}}  & 0 & 0 & 0 & 0 & 0 & 0 \\
0 & 0 & 0 & 0 & 0 & 0 & 0 & 10^{\text{-23}} &10^{\text{-21}}
& {\color{red}\mathbf{10^\textbf{-44}}} & 0 & 0 & 0 & 0 & 0 \\
0 & 0 & 0 & 0 & 0 & 0 & 0 & 0 &
{\color{red}\mathbf{10^\textbf{-44}}} &
10^{\text{-68}} & 0 & 0 & 0 & 0 & 0 \\
0 & 0 & 0 & 0 & 0 & 0 & 10^{\text{-10}} & 0 & 0 & 0 &
10^{\text{-9}} & 10^{\text{-17}} & 0 & 0 & 0 \\
0 & 0 & 0 & 0 & 0 & 0 & 0 & 0 & 0 & 0 & 10^{\text{-17}}
& 10^{\text{-16}} & 10^{\text{-33}} & 0 & 0 \\
0 & 0 & 0 & 0 & 0 & 0 & 0 & 0 & 0 & 0 & 0 & 10^{\text{-33}}
& 10^{\text{-31}} & 10^{\text{-65}} & 0 \\
0 & 0 & 0 & 0 & 0 & 0 & 0 & 0 & 0 & 0 & 0 & 0 & 10^{\text{-65}}
& 10^{\text{-61}} & {\color{red}\mathbf{10^\textbf{-130}}} \\
0 & 0 & 0 & 0 & 0 & 0 & 0 & 0 & 0 & 0 & 0 & 0 & 0 &
{\color{red}\mathbf{10^\textbf{-130}}} & 10^{\text{-197}}
\end{array}
\right)$}.

\noindent In this matrix, the ({\color{red}red}) boldface
non-diagonal elements represent the dimensionless coupling
constants of LE interactions. These elements really reproduce the
hierarchy experimentally known for coupling constants of the
fundamental interactions.\cite{EM} It is remarkable that in our
model the coupling constants of which $\Lambda^{ab}$ is built, are
basic topological invariants of the four-dimensional spacetime
$X$. In fact, we theoretically reproduced the vacuum coupling
constants hierarchy existing in the real universe if the spacetime
is a graph manifold. In our scheme the five LE interactions are
related to the first nine prime numbers. To obtain any new
interaction, one has to attach a new pair of prime numbers to the
preceding set. With the next pair (29,31), the same algorithm
yields a new coupling constant of the order of magnitude
$\alpha_6\approx 10^{-361}$. Thus our model answers the question:
How many fundamental interactions may exist in the universe? To
the infinite succession of prime numbers should correspond
infinite number of interactions. We simply cannot detect too weak
interactions beginning with $\alpha_6$, and all subsequent are
even much weaker.

\end{document}